\documentclass[aps,pra,reprint]{revtex4-2}

\usepackage[T1]{fontenc}
\usepackage{graphicx}
\usepackage{bbold}

\usepackage{enumitem}
\usepackage{float}
\usepackage[colorlinks=true,
            linkcolor=magenta,
            urlcolor=blue,
            citecolor=green]{hyperref}

\usepackage{amsmath, amsfonts, mathtools, amsthm, amssymb}
\usepackage{bm}
\usepackage{xcolor}

\DeclareMathOperator*{\sumint}{%
\mathchoice%
  {\ooalign{$\displaystyle\sum$\cr\hidewidth$\displaystyle\int$\hidewidth\cr}}
  {\ooalign{\raisebox{.14\height}{\scalebox{.7}{$\textstyle\sum$}}\cr\hidewidth$\textstyle\int$\hidewidth\cr}}
  {\ooalign{\raisebox{.2\height}{\scalebox{.6}{$\scriptstyle\sum$}}\cr$\scriptstyle\int$\cr}}
  {\ooalign{\raisebox{.2\height}{\scalebox{.6}{$\scriptstyle\sum$}}\cr$\scriptstyle\int$\cr}}
}

\def\probefreqData{1.53067}
\def\pumpfreqData{27.211386}
\def\IlowData{10^{11}\, \mathrm{W}/\mathrm{cm}^2}
\def\IhighData{10^{12}\, \mathrm{W}/\mathrm{cm}^2}

\def\tauData{35}
\def\truncationLengthData{33.88}
\def\toffTruncatedGaussian{4}
\def\tmaxTruncatedGaussian{6}
\def\tsurffRData{100\, \mathrm{Bohr}}

\begin{document}

\title{Thomas--Reiche--Kuhn Correction for Truncated Configuration Interaction Spaces: \texorpdfstring{\\C} Case of Laser-Assisted Dynamical Interference}
\affiliation{$^1$Affiliation}
\author{Mattias \surname{Bertolino}$^1$}
\author{Stefanos \surname{Carlstr\"om$^1$}}
\author{Jasper \surname{Peschel$^1$}}
\author{Felipe \surname{Zapata$^1$}}
\author{Eva \surname{Lindroth}$^2$}
\author{Jan~Marcus \surname{Dahlstr\"om}$^1$}
\affiliation{$^1$Department of Physics, Lund University, Box 118, SE-221 00 Lund, Sweden \\
$^2$Department of Physics, Stockholm University, AlbaNova University Center, SE-106 91}

\begin{abstract}
  The Thomas--Reiche--Kuhn sum rule is used to form an effective potential that is added to the time-dependent configuration interaction singles (TDCIS) equations of motion in velocity gauge.
  The purpose of the effective potential is to include virtual coupling from singles to doubles, which is required for size-consistent velocity gauge TDCIS results.
  The proposed method is compared to length gauge TDCIS results for laser-assisted photoionization. Finally, a dynamical interference effect controlled by two-color fields is predicted for atomic targets.
\end{abstract}

\maketitle

\section{Introduction}
\label{sec:introduction}
There are several frontiers of research in attosecond physics \cite{krauszAttosecondRMP2009}, including studies of charge migration in biorelevant molecules~\cite{calegariChargeJPBAMOP2016}, generation of attosecond pulses in solid-state targets~\cite{ghimireHighharmonicNP2019} and atomic delay measurements in laser-assisted photoionization \cite{isingerPhotoionizationS2017}.
It has been shown that attosecond precision measurements can be performed by photoelectron interferometry, using various forms of extreme ultraviolet (XUV) and infrared (IR) pulses, in atoms  \cite{paulObservationS2001,schultzeDelayS2010,laurentAttosecondPRL2012,marojuAttosecondN2020,isingerPhotoionizationS2017}, molecules \cite{haesslerPhaseresolvedPRA2009,wangRolePRA2021} and solids \cite{cavalieriAttosecondN2007,nepplDirectN2015}.
While this type of non-linear interferometry can be qualitatively understood using the Strong-Field Approximation (SFA),
as we have recently reviewed in Ref.~\cite{bertolinoMultiphotonPRR2021} for numerous experiments with light from High-order Harmonic Generation (HHG) and Free-Electron Laser (FEL) sources,
a more detailed interpretation requires use of advanced many-body theory.
%
%
In cases where the interaction with the fields is ``weak'' it is possible to employ perturbation theory to interpret attosecond delays from atoms ~\cite{dahlstromDiagrammaticPRA2012,vinbladhManybodyPRA2019},  molecules~\cite{baykushevaTheoryJCP2017,kamalovElectronPRA2020,bendaAnalysisPRA2022} and laser-stimulated transitions via autoionizing states~\cite{jimenez-galanModulationPRL2014,jimenez-galanTwophotonPRA2016,koturSpectralNC2016,grusonAttosecondS2016,argentiControlPRA2017,barreauDisentanglingPRL2019}.
However, when the interaction with fields increase in strength,  time-dependent simulations are essential to understand the correlation effects that accompany the laser-driven dynamics.
Brute force approaches for photoionization studies are possible only in two-electron systems, such as  He and H$^-$, where the Schrödinger equation can be directly propagated~\cite{parkerIntensefieldJPBAMOP1996,ossianderAttosecondNP2017}.
In other systems, time-dependent simulations of multi-electron dynamics are more costly and one must rely on approximations that balance precision against the numerical cost.
A few examples include the Time-Dependent Configuration Interaction Singles (TDCIS) \cite{rohringerConfigurationinteractionbasedPRA2006,greenmanImplementationPRA2010}, $R$-matrix ~\cite{mooreTimeJPCS2012,bendaPerturbativePRA2020}, X-CHEM ~\cite{marantePhotoionizationPRA2017,barreauDisentanglingPRL2019} and self-consistent field theories \cite{satoTimedependentPRA2015,miyagiTimedependentPRA2013,miyagiTimedependentPRA2014}. The latter theories rely on time-dependent Slater determinants, which form eigenstates of approximate time-dependent Hamiltonians. The simplest such example is the Time-Dependent Hartree Fock (TDHF) theory, where a single Slater determinant is constructed from time-dependent occupied orbitals.

In addition to the level of electron correlation included in the simulations, care must also be given to the question of gauges that describe the electromagnetic interaction with the electrons.
%
If no approximations were made,
the Schrödinger equation could have equivalently been expressed in either gauge.
However, the truncation of the basis makes the physical observables gauge-dependent~\cite{ishikawaReviewIJSTQE2015,rohringerConfigurationinteractionbasedPRA2006}. This raises the question of which gauge happens to give (i) the best correspondence to experiment, (ii) the most convenient numerical properties.
While the Hartree--Fock equations (HF) have been shown to be gauge invariant \cite{kobeGaugeinvariantPRA1979}, such a formulation requires the occupied orbitals to be gauge-transformed dynamically in the presence of time-dependent fields.
This implies that self-consistent field theories \cite{satoTimedependentPRA2015,miyagiTimedependentPRA2013,miyagiTimedependentPRA2014} are highly desirable in this regard, but, unfortunately, such theories are costly to propagate numerically.
According to Kobe's gauge theory \cite{kobeGaugeAJP1978}, the length gauge is unique with expansion coefficients that are genuine probability amplitudes.
This implies that dynamic gauge transformations are less essential, and that more efficient basis truncations can be performed in the length gauge, as compared to any other gauge.
However, the length gauge is notorious for being difficult to converge for strong-field processes with low-frequency laser pulses~\cite{reissTheoreticalPiQE1992,cormierOptimalJPBAMOP1996,mullerEfficientLP1999,schaferNumericalNmisfp2009}.
Indeed, a relatively low number of angular momenta is required for convergence of TDCIS \cite{rohringerConfigurationinteractionbasedPRA2006,greenmanImplementationPRA2010} in velocity gauge to describe laser-assisted photoionization~\cite{bertolinoPropensityJPBAMOP2020,bertolinoMultiphotonPRR2021}.
On the other hand, length gauge TDCIS calculations have empirically proven better than velocity gauge calculations~\cite{satoGaugeInvariantAS2018}, meaning that the converged TDCIS results in length gauge are more accurate than the converged velocity gauge results, in agreement with the gauge theory of Kobe~\cite{kobeGaugeAJP1978,kobeGaugeinvariantPRA1979}.
To overcome this gauge-problem in TDCIS, Sato et al.\ have developed a modification to TDCIS in velocity gauge by successively rotating the orbitals in each time step, effectively defining time-dependent orbitals, which are necessary for gauge invariance with the length gauge TDCIS results~\cite{wolfsbergDipoleJCP1955,kobeGaugeinvariantPRA1979,satoGaugeInvariantAS2018}.

In this article, we investigate the gauge-dependence problem in TDCIS, from a different perspective than Sato et al.~\cite{satoGaugeInvariantAS2018}, by bridging the space of single (S) excitations to that of virtual double (D) excitations.
This is done by systematically adjusting for the lack of ``core'' polarization of excited states in the TDCIS.
In Section~\ref{sec:theory}, a review of polarization effects in $N$-electron atoms is presented by usage of the Thomas--Reiche--Kuhn (TRK) sum rule, (see e.g.\ Ref.~\cite{kuhnUeberZP1925}).
We then introduce an effective potential of TRK-type to the equations of motion for velocity gauge TDCIS in Section~\ref{sec:method}.
In Section~\ref{sec:results}, results for laser-assisted photoionization are presented for velocity gauge TDCIS and the proposed TDCIS-TRK theory.
It is shown that TDCIS-TRK provides accurate results where velocity gauge TDCIS fails.
We also revisit dynamical interference in photoionization~\cite{toyotaSiegertstatePRA2007,bagheryEssentialPRL2017,jiangDynamicOEO2018}, and predict that the phenomenon is experimentally feasible in a new setting with two-color fields.
Finally, in Section~\ref{sec:conclusions} we conclude our findings.

\section{Theory}
\label{sec:theory}
An atom subject to an external laser field will have its dynamics ruled by the minimal coupling Hamiltonian
\begin{equation}
  \label{eq:H-minimal}
  \hat H(\mathbf{r}, t) = \hat H_{0} + \hat V_{1}(\mathbf{r}, t) + \hat V_{2}(\mathbf{r}, t),
\end{equation}
where the zeroth term $\hat{H}_0$ is the atomic Hamiltonian, which accounts for the kinetic energy and the Coulomb interaction with the nucleus and among the electrons.
For a wavelength substantially larger than the atom, $\lambda \gg a_0$, the spatial dependence of the field can be neglected.
This is referred to the dipole approximation, and when applied to the minimal coupling Hamiltonian, we refer to it as the \textit{velocity gauge}.
The interactions with the electromagnetic field are hence given by a term linear in the time-dependent vector potential:
\begin{equation}
  \label{eq:H-minimal-pA}
  \hat{V}_{1}(t) = -\frac{q}{m} \mathbf{A}(t) \cdot \sum_{ij} \mathbf{p}_{ij} \hat{c}_{i}^{\dagger} \hat{c}_{j},
\end{equation}
and a term quadratic in the vector potential:
\begin{equation}
  \label{eq:H-minimal-A2}
  \hat{V}_{2}(t) = \frac{q^{2}A^{2}(t)}{2m} \hat{n}.
\end{equation}
Here, $q=-e$ is the electronic charge and $\mathbf{p}_{ij} = \langle i|\hat{\mathbf{p}}|j\rangle$ denotes the matrix element for momentum between the canonical orbitals $i$ and $j$, which are eigenstates of the mean-field (Fock) operator $\hat{f}|i\rangle = \varepsilon_{i}|i\rangle$.
The operators $\hat{c}_{i}^{\dagger}$ and $\hat{c}_{i}$ creates respectively annihilates the canonical orbital $i$.
The $\hat V_2$ operator acts as a scalar, due to any $N$-body state being an eigenstates to the number operator,  $\hat{n} = \sum_{i} \hat{c}_{i}^{\dagger} \hat{c}_{i}$, such that $\hat n |\Psi\rangle=N|\Psi\rangle$.

\subsection{Energy shifts in an oscillating potential}

Consider a monochromatic vector potential $A(t) = A_{0}\cos(\omega t)$, polarized linearly along the $z$-axis.
The potential term linear in the vector potential is given by
\begin{equation}
\hat{V}_1(t)=\hat{V}_1^{(+)}\exp(-i\omega t) + \hat{V}_1^{(-)}\exp(i\omega t)
\end{equation}
with
\begin{equation}
    \hat{V}_{1}^{(+)} = \hat{V}_{1}^{(-)} = -\frac{q}{2m} A_{0} \sum_{ij} (p_{z})_{ij} \hat{c}_{i}^{\dagger} \hat{c}_{j},
\end{equation}
where the superscripts $(+)$ and $(-)$ correspond to interactions that induce absorption and emission of a photon, respectively.
The quadratic potential amplitude is given by
\begin{equation}
  \hat{V}_{2}(t) = \frac{q^2A_0^2\cos^2(\omega t)}{2m}\hat{n}.
\end{equation}

The second-order energy correction to any $N$-electron state, $|\Psi_0\rangle$, due to $\hat{V}_1(t)$ is given by
\begin{widetext}
  \begin{equation}
    \label{eq:energy-correction1}
    \langle \Delta E_{1}^{(2)}(\omega) \rangle = \lim_{\eta \to 0^+} \sumint_{n\neq 0} \left(\frac{\langle \Psi_0| \hat{V}_{1}^{(+)} | \Psi_{n}\rangle \langle  \Psi_{n} | \hat{V}_{1}^{(-)} | \Psi_{0} \rangle}{E_{0}  -\hslash\omega - E_{n} + i\eta} + \frac{\langle \Psi_0| \hat{V}_{1}^{(-)} | \Psi_{n}\rangle \langle  \Psi_{n} | \hat{V}_{1}^{(+)} | \Psi_{0} \rangle}{E_{0} +\hslash\omega- E_{n}  + i\eta}\right),
  \end{equation}
\end{widetext}
which can be interpreted as a level shift of the state averaged over time~\cite{sakuraiModern2017}.
The TRK sum rule states that the sum of all oscillator strengths, $f_{nn'}$, from a particular atomic eigenstate, $|\Psi_{n}\rangle$, is equal to the number of electrons:
\begin{equation}
  \label{eq:TRK-sum}
  \sum_{n'} f_{nn'} = \sum_{n'} \frac{2m}{\hslash^2} (E_{n} - E_{n'}) | \langle  \Psi_{n'} | \hat{z} | \Psi_{n} \rangle |^2 = N.
\end{equation}
The application of TRK theory to \eqref{eq:energy-correction1} requires that the commutation between the atomic Hamiltonian and the dipole operator equals the momentum operator, $[\hat{H_0}, \hat{z}] = -i\hslash \hat{p}_z/m$, (see e.g.\ Eq.~(61.1) in Ref.~\cite{betheQuantum1977}).
Since this condition is satisfied for any {\it local} potential, the TRK sum rule can be used to perform exact calculations in $N$-electron atoms, where all interactions are inherently local.
The time-averaged energy-shift is derived by Taylor expanding the denominators in Eq.~\eqref{eq:energy-correction1} with $\hslash\omega\ll|E_n-E_0|$ to yield an expression that is valid up to second order in $V_1(t)$:
\begin{equation}
  \label{eq:energy-correction1-a0}
  \langle \Delta E_{1}^{(2)}(\omega) \rangle = - \frac{A_{0}^{2}}{4}\left(\frac{q^2}{m}N + \omega^2\alpha_0 \right)  + \mathcal{O}\left(\omega^4\right),
\end{equation}
where
\begin{equation}
  \label{eq:polarizability}
  \alpha_0 = \frac{q^2 \hslash^2}{m}\sum_{n \neq 0} \frac{f_{n0}}{(E_0 - E_n)^2}
\end{equation}
is the atomic {\it polarizability}, as defined in Ref.~\cite{margenauVanPR1939}.
The time-averaged energy shift of the first-order energy correction due to the quadratic term, $\hat{V}_2(t)$, is simply given by
\begin{equation}
  \label{eq:energy-correction2}
  \langle \Delta E_{2}^{(1)}(\omega) \rangle = \frac{q^{2}\langle A^{2}(t) \rangle}{2m}N = \frac{q^{2} A_{0}^{2}}{4m} N.
\end{equation}
This shows that, in the case of a monochromatic field, the energy corrections due to $\hat{V}_{1}(t) + \hat{V}_{2}(t)$ to second order in the vector potential {\it cancel} such that the explicit $N$-dependence in the energy shift energy disappears. The remaining energy shift in velocity gauge is
\begin{equation}
  \label{eq:final-energy-correction}
  \langle \Delta E^{(2)}(\omega) \rangle = -\frac{\omega^2A_0^2\alpha_0}{4} + \mathcal{O} \left(\omega^4\right)=-\frac{E_0^2\alpha_0}{4}+ \mathcal{O}\left(\omega^4\right),
\end{equation}
which can be found more easily in the length gauge with electric field $E(t)=-\frac{dA}{dt}=E_0\sin(\omega t)$ by second-order perturbation theory with the dipole operator $\hat{V}_1^{(\mathrm{len.})}=-qE(t)\hat{z}$.
While length gauge is simpler in this regard, because it does not rely on a detailed cancellation effect of two different perturbation terms, the energy shifts are found to be equivalent in both gauges.
We stress that this exact result is general, as it applies to any state in atomic or molecular targets, but it fails in cases where the considered state resonates with another state at the applied frequency: $\hslash \omega\approx |E_n-E_0|$.
Thus, as pointed out by Bucksbaum et al.~\cite{bucksbaumRoleJOSABJ1987}, it is typically an excellent approximation for the ground state of an atom, or molecule, subject to a low-frequency laser field, $\hslash\omega\ll |E_n-E_0|$, but it is a bad approximation for excited states, where other Rydberg states are likely to be close in energy.
We note that the TRK rule has been used in solids to reduce the necessary number of energy bands in velocity gauge while ensuring convergence~\cite{yakovlevAdiabaticCPC2017}.
In the present work however, we will make use of the TRK sum rule in the numerical propagation to correct for size inconsistency in virtual excitations that are missing due to truncations of the many-body configuration-interaction expansion.

\subsection{Configuration Interaction (CI)}
Many-body effects can be described by Time-Dependent Configuration Interaction (TDCI) methods, where the total electron wavefunction is expanded as
\begin{equation}
  \label{eq:TDCI}
  | \Psi (t)\rangle = \alpha_0(t) | \Phi_0 \rangle + \sum_{ap} \alpha_a^p(t) | \Phi_a^p \rangle + \sum_{abpq} \alpha_{ab}^{pq}(t) | \Phi_{ab}^{pq} \rangle + ...,
\end{equation}
where $| \Phi_0 \rangle$ denotes the  Hartree--Fock (HF) reference Slater determinant, while  $|\Phi_{a}^{p}\rangle$ and $|\Phi_{ab}^{pq}\rangle$ denote single (S) and double (D) excitations, respectively.
The labels $a,b,...$ index the occupied orbitals, and the labels $p,q,...$ index the unoccupied orbitals in the initial reference state.
While the orbitals are found variationally, without the laser-action part of the Hamiltonian, the expansion coefficients $\alpha(t)$ are time-dependent quantities in TDCI methods.
The expansion in Eq.~\eqref{eq:TDCI} can be truncated in several ways.
If the expansion is truncated so that the CI space is limited to the level of single excitations it is known as  TDCIS~\cite{rohringerConfigurationinteractionbasedPRA2006,greenmanImplementationPRA2010}.
The singly excited states $| \Phi_{a}^{p} \rangle$ are in the non-relativistic case restricted to the spin-singlet state character constructed using second quantization as
\begin{equation}
  \label{eq:second-quant}
  | \Phi_{a}^{p} \rangle = \frac{1}{\sqrt{2}} \{\hat{c}_{p+}^{\dagger}\hat{c}_{a+} + \hat{c}_{p-}^{\dagger}\hat{c}_{a-} \} | \Phi_0 \rangle.
\end{equation}
The operator $\hat{c}_{p\sigma}^{\dagger}$ creates an electron in the virtual orbital $p$ with spin $\sigma$ and the operator $\hat{c}_{a\sigma}$ creates a hole in the core with orbital $a$ and spin $\sigma$.
While TDCI methods imply that occupied orbitals are static eigenstates of a laser-free mean-field Hamiltonian, electronic excitations at TDCIS-level can be interpreted as time-dependent wave packets in the form of superpositions of initially unoccupied orbitals~\cite{rohringerConfigurationinteractionbasedPRA2006}.
Recently, implementations of TDCIS that are not restricted to spin-singlet configurations have been realized, e.g.\ one in the form of two-component orbitals from relativistic pseudopotentials~\cite{carlstromGeneral2022,carlstromGeneralAP2022}, and another based on the four-component orbitals from the Dirac--Fock equation~\cite{zapataRelativisticPRA2022}.
In the following work we will consider only the spin-singlet configurations and linearly polarized fields, such that the magnetic quantum number is conserved: $m_p=m_a$, where in addition the {\it gerade} symmetry of TDCIS can be utilized to further reduce the number of correlated channels~\cite{pabstImpactPRA2012}.
Further truncation in CI can be performed to limit the allowed excitations, e.g.\ ``freezing'' occupied orbitals corresponding to tightly bound electrons
cf.\ Refs.~\cite{olsenDeterminantJCP1988,hochstuhlTimedependentPRA2012,satoTimedependentPRA2015,miyagiTimedependentPRA2013,miyagiTimedependentPRA2014}.

\subsection{Application of TRK to TDCIS}

Truncation of the Hilbert space implies that the TRK sum rules will not hold. %
Nonetheless, we consider the CI space to be truncated at the level of single excitations (CIS), where the states in Eq.~\eqref{eq:energy-correction1} are given by the space spanned by the HF ground state and the CIS singly excited states, $\{|\Psi_{0} \rangle, |\Psi_{n} \rangle\} \to \{|\Phi_{0}\rangle, | \Phi_{a}^{p} \rangle\}$, respectively.
We denote the space of no excitations by $\mathcal{P}_{0}$, single excitations by $\mathcal{P}_{1}$ and double excitations by $\mathcal{P}_{2}$.
\begin{figure*}[ht]
  \centering
  \includegraphics[width=\textwidth]{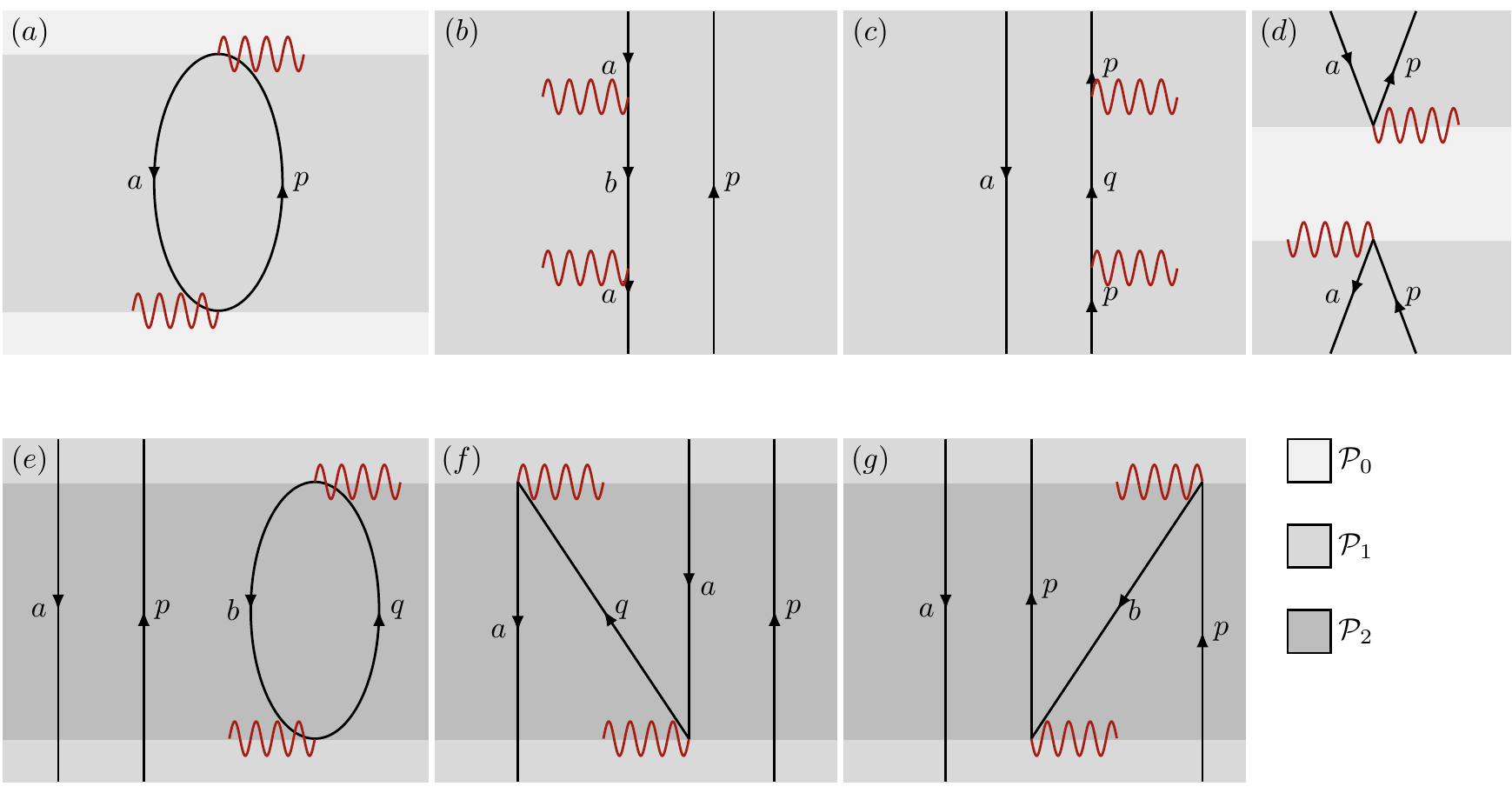}
  \caption{Goldstone diagrams of the energy contributions within (a--d) CIS and (e--g) CID.
  The diagram in (a) describes the energy shift of the core, while the diagrams in (b--g) describe the energy shift of an excited CIS state.
  The background shading color indicates the CI space where the atom resides. Time flows upwards in these Goldstone diagrams.}
  \label{fig:goldstone}
\end{figure*}
In Figure~\ref{fig:goldstone}, we illustrate some processes from perturbation theory, represented as Goldstone diagrams, that give rise to energy shifts due to second-order action of $\hat{V}_1(t)$.
The first diagram: Fig.~\ref{fig:goldstone}~(a), can be interpreted as an uncorrelated (bare) energy correction of the HF ground state, $|\Phi_0\rangle$ in $\mathcal{P}_{0}$, due to a virtual transition via the singly excited state, $|\Phi_a^p\rangle$ in $\mathcal{P}_1$.
Since in TDCIS theory both $\mathcal{P}_0$ and $\mathcal{P}_1$ are contained in the truncated Hilbert space, the polarization of the HF ground state will be reasonably well described.
The singly excited states can also shift their energies due to second-order action with $V_1(t)$ within $\mathcal{P}_1$ and via $\mathcal{P}_0$.
These effects are described in TDCIS theory, as shown in Fig.~\ref{fig:goldstone}~(b--d).
However, some perturbative processes that also change the energy of the singly excited states are missing, since they require a virtual transition via the \textit{doubly} excited states in $\mathcal{P}_2$, as shown in Fig.~\ref{fig:goldstone}~(e--g).
The former diagram: Fig.~\ref{fig:goldstone}~(e), shows that the laser field can polarize the core of the atom, similar to how the ground state was excited: Fig.~\ref{fig:goldstone}~(a), which leads to an energy shift $\sim N$, in accordance with Eq.~\eqref{eq:energy-correction1-a0}.
However, the atomic core is missing one-electron \textit{exchange} processes: Fig.~\ref{fig:goldstone}~(f,g) that impose the Pauli exclusion principle, which will reduce the core energy shift to $\sim N-1$.
Additionally, the reversed time order of Fig.~\ref{fig:goldstone}~(d) is also missing in CIS and further many-body polarization effects, induced by Coulomb interactions, must be carefully considered by Many-Body Perturbation Theory (MBPT) for accurate energy shifts.

Since the diagrams in Fig.~\ref{fig:goldstone}(e--g) are non-existent in TDCIS, the time evolution will be subject to a spurious effects.
While the polarization effects are reasonably accounted for in the ground state, the excited states in CIS have an unbalanced polarization that lacks size consistency.
Since transitions between the ground state and the excited states are essential in TDCIS theory this inconsistency will result in \textit{unphysical} relative energy shifts for excitations.
These type of relative energy shifts are usually avoided in MBPT, by the linked diagram theorem, (see e.g.\ Ref.~\cite{lindgrenAtomic1986}), but in time-dependent simulations, such as TDCIS, the problem must be treated in a different way.
We will introduce an effective potential that corrects the relative energy shifts based on TRK theory.
In order to understand this better we first consider the energy shifts computed for the HF ground state of neon with various approximations using MBPT.

Due to the non-locality of the Hartree--Fock exchange potential, when static or frozen occupied orbitals are used, the second-order energy correction is not exactly given by Eq.~\eqref{eq:energy-correction1-a0}.
Instead, we define the ground state energy shift
\begin{equation}
  \label{eq:CIS-energy-correction1}
  \langle \Delta \tilde E_{1}^{(2)} (\omega) \rangle = -\frac{A_{0}^{2}}{4} \left( \frac{q^2}{m}\tilde{N} + \omega^2 \tilde\alpha_0 \right) + \mathcal{O} \left(\omega^4\right),
\end{equation}
where $\tilde{N}$ and $\tilde \alpha_0$ denote the corresponding value of $N$ and $\alpha_0$ for a given approximation: Lowest-Order Perturbation (LOP), Configuration Interaction Singles (CIS) and Random Phase Approximation with Exchange (RPAE).
The {\it effective} number of active electrons, which is a frequency independent quantity, is computed as
\begin{equation}
  \label{eq:approximate-N}
  \tilde{N} \equiv
  -\left(\frac{8m}{A_0^2q^2}\right)
  \sum_{a}^\mathrm{occ.} \langle a | \hat{V}_1 | \rho_{0,a}^{(+)} \rangle,
\end{equation}
where $a$ label all occupied orbitals in from the HF ground state and $|\rho_{0,a}^{(+)}\rangle$ is an associated perturbed wavefunction that depends on the level of approximation.
In the case of RPAE, the perturbed wavefunctions are (in atomic units) constructed as~\cite{amusiaAtomic1990}
\begin{widetext}
\begin{equation}
  \label{eq:rho-a}
    | \rho_{\omega,a}^{(\pm)} \rangle = \sum_{p}^{\mathrm{exc.}} \frac{|p\rangle}{\varepsilon_{a} - \varepsilon_{p} \pm \omega}
    \bigg{[} \langle p | \hat{V}_1 | a \rangle -
    \sum_{b}^{\mathrm{occ.}} \bigg{(}
    \langle b,p | r_{12}^{-1} | a, \rho_{\omega,b}^{(\pm)} \rangle
    -
    2
    \langle b,p | r_{12}^{-1} | \rho_{\omega,b}^{(\pm)},a \rangle
    +
    \langle  \rho_{\omega,b}^{(\mp)},p | r_{12}^{-1} | a,b \rangle
    -
    2
    \langle  \rho_{\omega,b}^{(\mp)},p | r_{12}^{-1} | b,a \rangle \bigg{)}\bigg{]},
\end{equation}
\end{widetext}
where the superscripts $(+)$ and $(-)$ denote forward- and backward-propagating perturbed wavefunctions.
Note that it is the $(+)$-function at zero-frequency, $\omega=0$, that is inserted into Eq.~\eqref{eq:approximate-N}.
The first $(+)$-term on the right-hand side of Eq.~\eqref{eq:rho-a} corresponds to uncorrelated (bare) excitation, which we label as LOP.
Including also the second and third $(+)$-terms correspond to forward-propagating electron--hole correlations, which we label CIS, because it is the level of correlation obtained by solving TDCIS (also known as Tamm--Dancoff, (see e.g.\cite{amusiaAtomic1990}).
The fourth and fifth terms are exclusive to RPAE and they correspond to ground state correlation effects (direct and exchange) by self-consistent solution of both $(+)$ and $(-)$-terms.
More details about the use of perturbed wavefunctions in attosecond physics, and how they correspond to dressing of the lower vertex in Fig.~\ref{fig:goldstone}~(a), are found in Refs.~\cite{dahlstromDiagrammaticPRA2012,vinbladhManybodyPRA2019}.
In Table~\ref{tab:N-values} the numerically obtained values of $\tilde{N}$ for LOP, CIS, and RPAE are presented together with the number of active electrons in the core of neon atoms.
RPAE is exceptional since it corresponds to the linear response of the TDHF approximation, which is a self-consistent field theory, and therefore does obtain the correct number of electrons.
However, when the active core is truncated to $N_\mathrm{A}<N$, RPAE does {\it not} provide the correct number of active electrons.
It is interesting to note that when only $2p$ orbitals are active, the CIS approximation outperforms the RPAE approximation in this aspect.
\begin{table}[H]
   \centering
   \caption{Effective number of active electrons $\tilde{N}$ within LOP, CIS, and RPAE, for the neon atom $(1s^22s^22p^6)$ given the number of electrons in the  active core: $N_\mathrm{A}$.}
   \label{tab:N-values}
  \begin{tabular}{rrrrr}
	 \hline
  Active & $\tilde{N}_{\mathrm{LOP}}$ & $\tilde{N}_{\mathrm{CIS}}$ & $\tilde{N}_{\mathrm{RPAE}}$ & $N_{\mathrm{A}}$ \\
	 \hline
  2p & 5.4091 & 6.1758 & 7.2461 & 6 \\
  2s,2p & 6.2712 & 7.2558 & 8.3022 & 8 \\
  1s,2s,2p & 7.8528 & 8.8858 & 10.0000 & 10 \\
	 \hline
\end{tabular}

\end{table}

\subsection{Energy shifts in a static potential}

Alternatively, static energy shifts of the HF ground state, $|\Phi_0\rangle$, in response to an external potential, $V_1$, can be obtained by diagonalizing the Hamiltonian, $\hat H_0+\hat V_1$ expressed within CIS.
In this case  a constant static vector potential, $A^{(\mathrm{stat.})}=A_0$, in the linear interaction term in Eq.~\eqref{eq:H-minimal-pA} is used.
Since the action of the quadratic term is trivial in this case it is omitted in the following discussion.
This approach has the advantage that it is not restricted to the second-order interaction, but includes all higher order corrections with the field directly.
Diagonalization of the CIS Hamiltonian can also be used to study non-linear DC Stark shifts of excited (Rydberg) states.
This feature is however of limited use since laser fields with finite frequencies will induce AC Stark shifts that differ significantly from the DC results due to resonances in the Rydberg series.

\begin{figure}[H]
  \centering
  \includegraphics[width=0.49\textwidth]{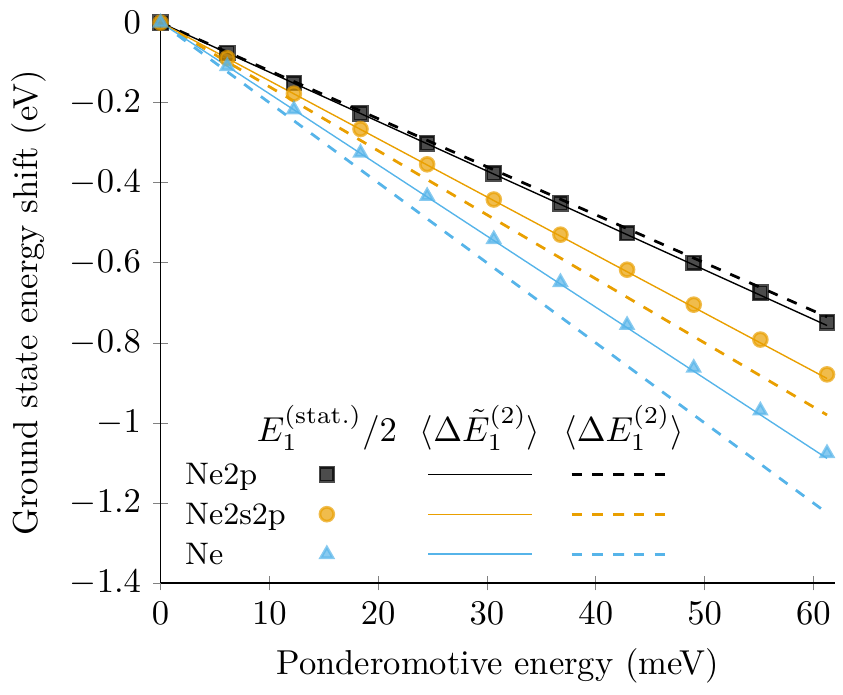}
  \caption{Calculated polarization of the HF ground state of neon with an applied static vector potential in velocity gauge with truncated space and full CIS space. Perturbation theory results are included as guiding lines.}
  \label{fig:polarization}
\end{figure}

In Figure~\ref{fig:polarization}, we plot {\it half} static energy shifts, $E_{1}^\mathrm{(stat.)}/2$, as a function the ponderomotive energy of an electron, $U_p = \frac{q^2 A_0^2}{4m}$, in an oscillating vector potential with the corresponding amplitude, $A^{(\mathrm{osci.})}(t)=A_0 \cos(\omega t)$.
The half static energy shifts of the HF ground state follow the time-averaged energy shifts, predicted by Eq.~\eqref{eq:energy-correction1-a0} with $N$ substituted by $\tilde{N}_{\mathrm{CIS}}$ from Table.~\ref{tab:N-values}, as shown by solid lines, $-\tilde N_\mathrm{CIS} U_p$.
There is a factor of two difference between the static energy shifts and those predicted by Eq.~\eqref{eq:energy-correction1} for monochromatic pulses because energy conservation is respected only by acting once with $V^{(-)}$ and once with $V^{(+)}$, in any time order, but not with twice the action of the same $V^{(\pm)}$.
In the limit $\omega \to 0$, however, conservation of energy is fulfilled for all four cases, yielding a factor of two compared to the time-averaged energy shift,
\begin{equation}
\Delta E_{1}^{(\mathrm{stat.})} \approx 2 \left<\Delta \tilde E_{1}^{(2)}(\omega\rightarrow 0)\right>.
\end{equation}
At high ponderomotive energies, higher-order interactions with the static field lead to a slight disagreement with the perturbative results, $-\tilde N_\mathrm{CIS}U_p$, as can be observed in Fig.~\ref{fig:polarization}.
Energy shifts equal to $-N_\mathrm{A}U_p$ are shown as dashed lines for reference.
 The significant discrepancy between the diagonalized ground state energy shift in CIS and the exact TRK theory, $-N_{\mathrm{A}}U_p$, originates from the truncations performed in the CI expansion, which leads to a non-local potential in TDCIS theory.

\section{Method}
\label{sec:method}
\subsection{Effective potential}

Effective potentials have found several important applications in atomic, molecular, and optical physics.
The key point is to reduce the complexity of a problem by replacing some degrees of freedom with a potential that depends on energy.
Effective potentials can be employed to perform {\it self-energy correction} to holes,  due to virtual Auger or shake-up processes, or, more generally, to include virtual many-body interactions into a subspace $\mathcal{P}$ from a complement space $\mathcal{Q}$ ~\cite{lindgrenAtomic1986,amusiaAtomic1990}.
Similarly, \textit{optical potentials} are used in scattering theory to include virtual processes in the target that are induced by the impinging particle   \cite{friedrichTheoretical1991}.
Effective potentials can also be used to include laser dressing effects, such as AC Stark shifts into a two-level system ($\mathcal{P}$), due to virtual interactions with the complement of the truncated Hilbert space ($\mathcal{Q}$) ~\cite{beersExactPRA1975,holtTimePRA1983}.
In this work, we introduce an effective potential that corrects for the lack of core polarization effects in the velocity gauge formulation of TDCIS theory due to virtual coupling to double excitations in the time domain.

We make the following assumptions for the core of $N-1$ electrons that remain after a single excitation from the ground state:

{\it 1. Markovian process:} The excitations from singles (S) to doubles (D) are virtual processes that only depend on the instantaneous value of the vector potential: $A(t)$, with no lasting memory effects in the state of the core over time.

{\it 2. Non-resonant dynamics:} The energy shift of the core is described by non-resonant dynamics that can be evaluated by TRK theory as: $-(N-1)q^2A^2(t)/2m$, at each given time, $t$, by virtual coupling from $\mathcal{P}=\mathcal{P}_1$ to $\mathcal{Q}=\mathcal{P}_2$.

{\it 3. Non-local correspondence:} The effect of exchange interactions in TDCIS are adopted  into the core energy shift using the substitution: $N-1 \rightarrow \tilde N-1$, where $\tilde N$ is the effective number of active electrons in the HF ground state.

{\it 4. Uniform action:} The core polarization energy affects all singly exited states in the same way independent of their individual energy and symmetry.

Using this set of assumptions, we present an effective potential for singles (S):
\begin{align}
\label{eq:vtrk}
    \hat V_\mathrm{TRK}^{\mathrm{(S)}}(t)={\mathcal V}_\mathrm{TRK}(t)\hat P_1=-\frac{q^2}{2m}(\tilde N-1)A^2(t)\hat P_1,
\end{align}
where
\begin{equation}
  \label{eq:Pq-projector}
  \hat P_1=\sum_p^\mathrm{virt.}\sum_a^\mathrm{occ.}|\Phi_a^p\rangle \langle \Phi_a^p|,
\end{equation}
is a $\mathcal{P}_1$-projector that acts on the singly excited states.
The effective potential in Eq.~\eqref{eq:vtrk} could be further improved, by adding the polarization energy of the core, $-\alpha_0^{(N-1)}E^2(t)/2$, from Eq.~\eqref{eq:final-energy-correction} for the $N-1$ subsystem.
We have {\it not} done this because that type of correction requires knowledge of the core system at a level that is {\it beyond} the CIS framework.
We stress that such core polarization effects are also missing in the length gauge formulation of TDCIS [see Fig.~\ref{fig:goldstone}~(e)].
We mention that it is straightforward to widen the present concept for generation of effective potentials that correct for virtual triples from doubles:
$\hat{V}_\mathrm{TRK}^\mathrm{(D)}(t)$ etc., but since such potentials are beyond TDCIS they are not considered in the present work.

\subsection{Equations of motion}
We define the CIS-TRK Hamiltonian as
\begin{align}
\label{eq:ham}
\hat{H}_{\mathrm{TRK}} &\equiv
\hat{H}_{\mathrm{CIS}}+\hat V^{\mathrm{(S)}}_{\mathrm{TRK}},
\end{align}
where $\hat H_\mathrm{CIS}$ is the original CIS Hamiltonian for velocity gauge ~\cite{greenmanImplementationPRA2010,karamatskouCalculationPRA2014}.
Due to the closure of the CIS-space, $\hat I_\mathrm{CIS}=\hat P_0+\hat P_1$, the effective potential can be moved to instead act exclusively on ground state. This is done by subtracting a time-dependent term from the CIS-TRK Hamiltonian:
\begin{align}
\hat H'_\mathrm{TRK} \equiv \hat{H}_\mathrm{TRK} - {\mathcal V}_\mathrm{TRK}(t)\hat I_{\mathrm{CIS}}
&=\hat H_{\mathrm{CIS}} - {\mathcal V}_\mathrm{TRK}(t)\hat P_0, \nonumber
\end{align}
where $P_0=|\Phi_0\rangle \langle \Phi_0|$.
This substitution, $\hat H_\mathrm{TRK}\rightarrow \hat H_\mathrm{TRK}'$, can formally be interpreted as a time-dependent gauge transformation that will modify the phase of the wavefunctions, but not alter any physical observables generated from the CIS-TRK theory.

The TDCIS-TRK$'$ equations of motions are (in atomic units) written as
\begin{widetext}
\begin{equation}
  \label{eq:tdcis}
  \begin{split}
    i \dot{\alpha}_0(t) &= \sqrt{2} A(t) \sum_{ap} \langle a|\hat{p}_{z}|p \rangle \alpha_{a}^{p}(t) + \frac{\tilde{N}-1}{2}A^2(t) \alpha_{0}(t) \\
    i \dot{\alpha}_{a}^{p}(t) &= (\varepsilon_{p} - \varepsilon_{a})\alpha_{a}^{p}(t)
    + \sum_{bq} [ 2 \langle bp | r_{12}^{-1} | qa\rangle -  \langle  bp | r_{12}^{-1} | aq \rangle ] \alpha_{b}^{q}(t) \\
    &+ A(t) \bigg{(} \sqrt{2} \langle p | \hat{p}_z | a \rangle \alpha_0(t) + \sum_{q} \langle  p | \hat{p}_z | q \rangle \alpha_{a}^{q}(t)
    -\sum_{b} \langle b | \hat{p} | a \rangle \alpha_{b}^{p}(t) \bigg{)},
  \end{split}
\end{equation}
\end{widetext}
which differ only by the addition of the effective potential in the ground state amplitude equation, when compared with the original TDCIS formulations  ~\cite{greenmanImplementationPRA2010,karamatskouCalculationPRA2014}.
In writing Eq.~\eqref{eq:tdcis} we have omitted the $\hat V_2(t)$-operator, because it will affect {\it all} CIS states in the same way within the dipole approximation.
As a result,  it is possible to remove it without affecting any physical observable generated from the theory.
We stress that the cost of implementing the TDCIS-TRK$'$ theory in Eq.~\eqref{eq:tdcis} is only one {\it scalar} multiplication per time step in the numerical propagation.
In Section~\ref{sec:results}, we will show that this seemingly minor correction to the equations of motion have important implications for the TDCIS theory.

\section{Results}
\label{sec:results}

In order to validate the action of the effective potential in Eq.~\eqref{eq:vtrk}, we perform \textit{truncated} TDCIS simulations for laser-assisted photoionization for neon atoms with realistic pulse parameters corresponding to an XUV pulse and an IR laser pulse, as defined in Eq.~\ref{eq:fields}.
The pulse duration is set to $\tau = \tauData{}\, \mathrm{fs}$, which is a typical value for XUV pulses generated by HHG or FEL.
The properties of the IR field are chosen to correspond to a Ti:Sapph.\ laser system.
In general, the photoelectron peak position depends on the detailed pulse forms of both the IR and XUV fields in laser-assisted photoionization.
This is because the ponderomotive energy,
\begin{equation}
\label{eq:up}
U_p(t) \equiv U_p^0 f^2(t) \approx q^2{\mathcal A}_{\omega,0}^2(t)/4m,
\end{equation}
with ${\mathcal{A}}_{\omega,0}(t)=A_{\omega,0}f(t)$, is dominated by the IR field, while the probability of one-photon ionization by the central XUV photon, $\Omega$, depends on time through the squared envelope of the XUV field:
\begin{equation}
\label{eq:p}
\dot{P} \sim |{\mathcal A}_{\Omega,0}(t)|^2=|{A}_{\Omega,0}f(t)|^2.
\end{equation}
In this way, the two fields have clearly defined roles: the IR field controls the dressing of the continuum, while the XUV field controls the flux of ejected electrons.

\subsection{Energy shift of the photoelectron}

\begin{figure*}[ht]
  \centering
  \includegraphics[width=\textwidth]{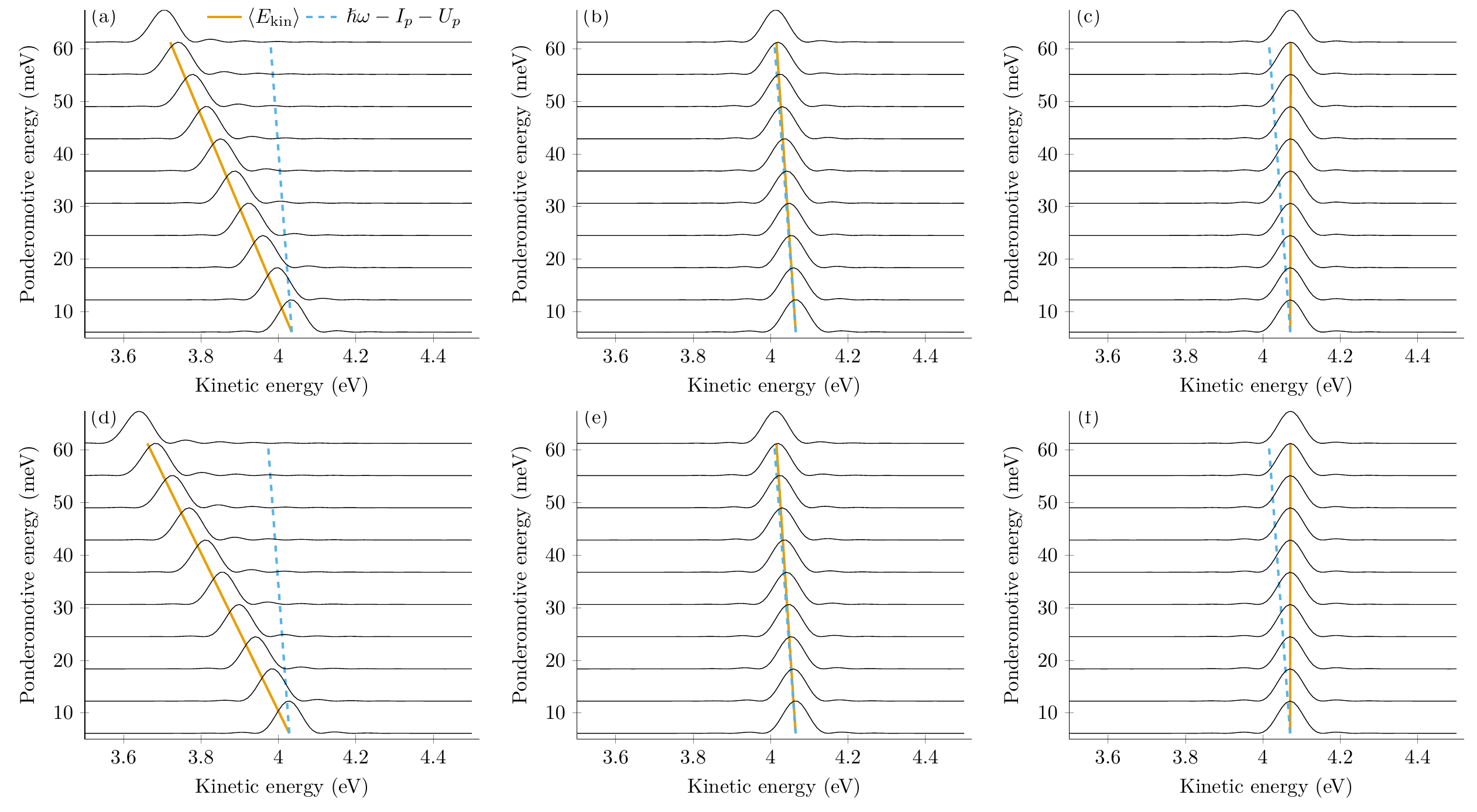}
  \caption{Normalized photoionization spectra using velocity gauge TDCIS in neon with (a--c) the 2p-orbital active and (d--f) the 2s- and 2p-orbitals active.
    The solid lines show measured peak position and the dashed lines show the expected peak position given by $\Delta E_{\mathrm{kin}} = -U_p$.
    The middle column (b,e) displays velocity gauge TDCIS-TRK with the effective potential $\hat{V}_{\mathrm{TRK}}^{(\mathrm{S})}$ and the rightmost column (c,f) displays velocity gauge TDCIS-TRK with the substitution $\tilde{N}-1 \to \tilde{N}$.}
  \label{fig:peaks}
\end{figure*}

In this subsection we consider the case where both the XUV and IR pulses have flat-top envelopes as defined in Eq.~\eqref{eq:flat-top}.
In Figure~\ref{fig:peaks}~(a), we show photoelectron probability distributions for absorption of one XUV photon from the neon ground state. Truncated TDCIS theory is used with only the $2p$ orbital active.
The intensity of the IR field is kept in the range $I=\IlowData{}$ to $I=\IhighData{}$, such that its main consequence is to assist the XUV photoionization process with the formation of sidebands (not shown), but also to shift the photoelectron structure to lower kinetic energies due to an increased ponderomotive potential \cite{bucksbaumRoleJOSABJ1987}.
A classical estimate for the kinetic energy of the photoelectron is given by:  $\left<E_\mathrm{\mathrm{kin}}\right> \approx \hslash \Omega + \varepsilon_{2p} - U_p^0$  (dashed sky blue line), where $\varepsilon_{2p}<0$ is the energy of the $2p$ orbital.
As will be shown, this  estimate is valid provided that both IR and XUV fields are flat-top pulses of the same duration, but with truncated TDCIS we instead observe a shift of the peak position of roughly $E_{\mathrm{kin}} \approx \hslash \omega + \varepsilon_{2p} - N_{\mathrm{A}}U_{p}^{0}$ (solid orange line), where $N_{\mathrm{A}} \approx 6$ because only the $2p$ orbital is open.
Obviously, this TDCIS shift is much too large when compared to the classical estimate.
In order to understand the convergence properties of TDCIS, we expand the active space to include both $2s$ and $2p$ orbitals.
This leads to an even stronger effect, as shown in Fig.~\ref{fig:peaks}~(d), and thus worse agreement with the classical estimate.
Finally, in the case where all orbitals are active, we find that the agreement with the classical prediction is further worsened (not shown).
This proves that extending the active core of TDCIS to {\it full} TDCIS does not resolve the issue, but rather confirms that the major issue with velocity gauge TDCIS is a synthetic AC Stark shift due to size inconsistency.

In Fig.~\ref{fig:peaks}~(b) and (e), we show the results of TDCIS-TRK theory, with the equation of motion defined in Eq.~\eqref{eq:tdcis}, for the truncated  active cores: $\{2p\}$ and $\{2s,2p\}$, respectively.
It is observed that the photoelectron peak for single XUV absorption follows the expected classical prediction closely (dashed line).
Furthermore, the two sets of simulations yield graphically equivalent results, which implies that the addition of the active $2s$ orbital was not essential to describe the physical process.

In order to study the role of the effective potential further, we make the substitution: $\tilde N-1\rightarrow \tilde N$ in Eq.~\ref{eq:tdcis}, and present the corresponding photoelectron peaks in Fig.~\ref{fig:peaks}~(c) and (f), with active cores, $\{2p\}$ and $\{2s,2p\}$, respectively.
Quite remarkably, it is found that the photoelectron peak now remains fixed at the same kinetic energy independent of the laser intensity.
This (unphysical) substitution corresponds to neglecting that one electron is removed from the atom in the process of photoionization and it shows that the exact value of effective number of active electrons, $\tilde{N}$ in Table~\ref{tab:N-values}, is crucial to understand the ponderomotive shift of photoelectrons.
In this way we have verified that the assumption of {\it non-local correspondence} between the $(N-1)$-body ionic core and the $N$-body atom is correct.

\subsection{Comparison with length gauge}
\begin{figure}[ht]
  \centering
  \includegraphics[width=0.49\textwidth]{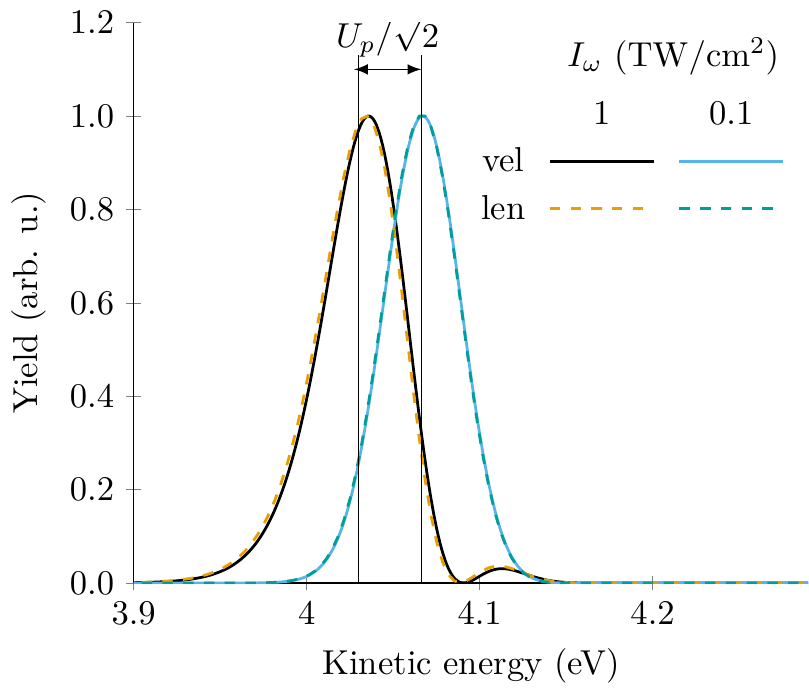}
  \caption{One-photon peak of laser-assisted photoionization with a truncated Gaussian envelope using both TDCIS-TRK velocity gauge and length gauge.
    Expected classical energy shift for a Gaussian envelope is shown with an arrow labeled $\Delta U_p/\sqrt{2}$.}
  \label{fig:lg-vg-tgau}
\end{figure}

In this subsection we consider the case where the pulses have more realistic time-dependent  envelopes, specifically given by the truncated Gaussian envelopes defined in Eq.~\eqref{eq:gaussian}.
Photoelectron probability distributions for one-photon ionization by an XUV field with an assisting laser field are shown using TDCIS-TRK theory in Figure~\ref{fig:lg-vg-tgau}.
A classical estimate of the mean kinetic energy shift of the photoelectron gives: $\left<\Delta E_\mathrm{kin}\right>\approx -\Delta U_p^0/\sqrt{2}$, provided that both IR and XUV fields are Gaussian pulses of the same duration.
The classical estimate for the photoelectron shift is marked with an arrow and shows good agreement with the numerical simulations.
As expected from the classical estimates, the photoelectron peak shifts less when Gaussian pulses, instead of flat-top pulses, are employed.
The quantum mechanical result shows a slightly smaller shift than the classical Gaussian estimate.
Finally, we have verified that the photoelectron spectra from length gauge TDCIS simulations (dashed lines) are in excellent agreement with the corresponding TRK results for the normalized yield.
On the blue side of the high-intensity peak ($I_{\omega}=\IhighData$), we observe an interference ripple in both gauges. We propose that this phenomenon is the onset of dynamical interference due to ionization at the rising and the falling sides of the field, which is discussed in more detail in the following section.

\subsection{Laser-assisted dynamical interference}

\begin{figure*}[t]
    \centering
    \includegraphics[width=\textwidth]{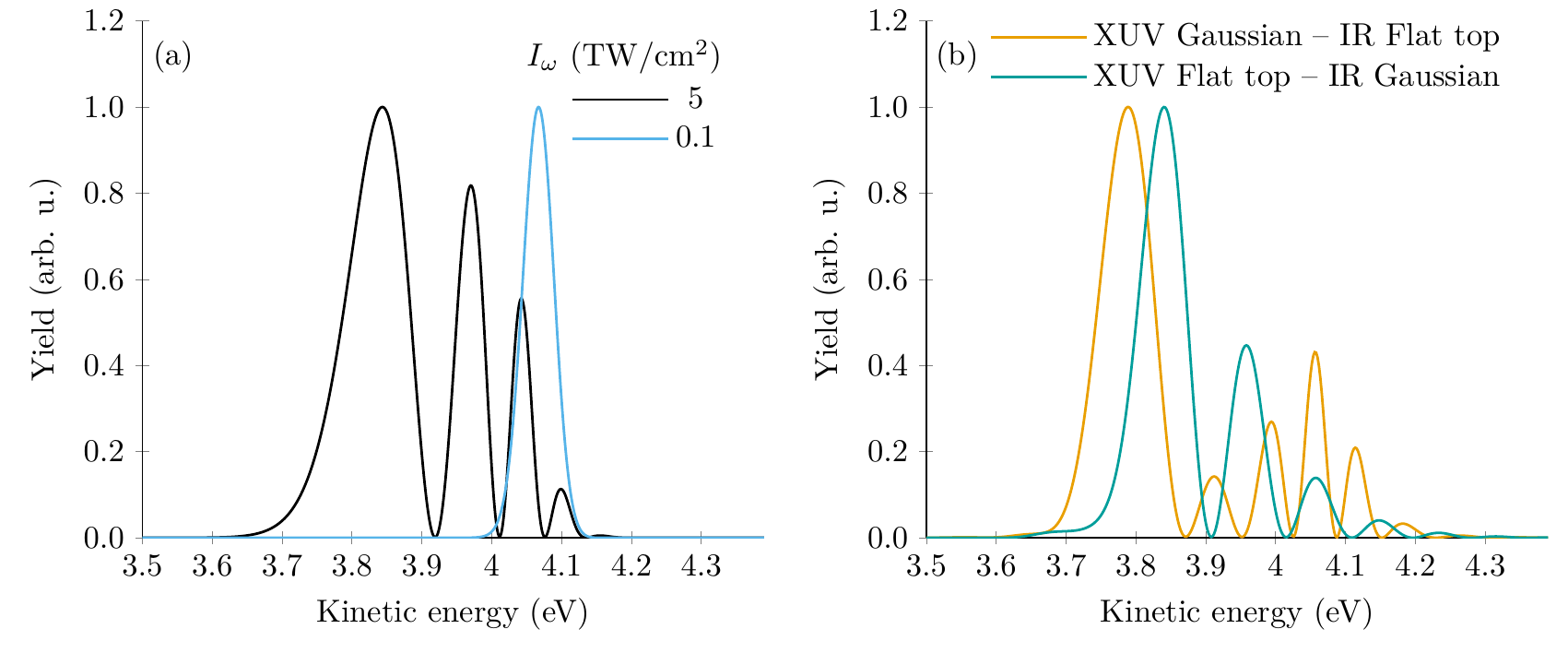}
    \caption{Demonstration of laser-assisted dynamical interference numerically using a single XUV pulse by
      (a) varying intensity of the IR field, and
      (b) mixing flat-top and truncated Gaussian envelopes of the XUV and IR fields.}
    \label{fig:ladi}
\end{figure*}

In Ref.~\cite{bagheryEssentialPRL2017} two criteria to observe dynamical interference are put forward: (i) the relative AC Stark shift between the initial and final states need to be larger than the bandwidth of the pulse and (ii) the ionization rate should not be so large that the initial state is depleted on the rising side of the field.
In practice this makes the experimental observation of dynamical interference very challenging, as it must rely on atomic stabilization mechanisms, due to the inherently small ponderomotive shifts of short-wavelength radiation~\cite{toyotaSiegertstatePRA2007,jiangDynamicOEO2018}.
Here, we propose that the usage of two-color fields, composed of XUV and IR parts, make the study of dynamical interference experimentally feasible with XUV fields from HHG or FEL sources.
The reason for this is that the flux and ponderomotive shift of photoelectrons are not constrained by a single field, as previously considered in Refs.~\cite{toyotaSiegertstatePRA2007,bagheryEssentialPRL2017,jiangDynamicOEO2018,dellapiccaNonconstantPRA2016}, but that they can be separately controlled by tuning the IR and XUV fields independently.
In Figure~\ref{fig:ladi}~(a) we show that dynamical interference is observed when the IR intensity is increased from $1\times10^{11}$ to $5\times 10^{12}$ W/cm$^2$.

The laser-assisted dynamical interference pattern can be studied further by choosing two different  envelopes for the XUV and IR pulses, as shown  in Fig.~\ref{fig:ladi}~(b), with a peak IR intensity of $5\times10^{12}$ W/cm$^2$.
The dramatic change of the interference pattern is the result of the different ponderomotive potentials of the IR field and the time-dependent photoionization triggered by the XUV field.
The case with a flat-top XUV and Gaussian IR field, shown in green in Fig.~\ref{fig:ladi}~(b), gives an interference pattern that is rather similar to the usual dynamical interference phenomenon, shown in Fig.~\ref{fig:ladi}~(a), with peaks on the blue side that decrease monotonically in magnitude and increase in oscillation rate with kinetic energy.
The case with a Gaussian XUV and flat-top IR field, shown in yellow in Fig.~\ref{fig:ladi}~(b), displays a more significant shift of the main peak, as expected for flat-top IR fields using classical arguments.
The interference fringes on the blue side, however, exhibit a qualitatively different behaviour with non-monotonic peaks.
We interpret this effect as Ramsey-like fringes from unshifted photoemission by the XUV pulse before and after the flat-top IR field is present.
The observed Ramsey fringes are $\hbar\Delta \omega_{\mathrm{R}} \approx 60$ meV, given $2\pi=\Delta \omega_{\mathrm{R}} T$ this implies a time separation of $T \approx 64$ fs.
We note that this separation is larger than the pulse duration of the flat-top IR field ($35$ fs), by almost a factor of two, which we interpret as a result of the temporal spread of the XUV Gaussian flank distributions (from $\pm 17.5$ fs to $\pm \infty$) with mean values at $\pm 29$ fs.

\section{Conclusions}
\label{sec:conclusions}
In this article, we discussed the velocity-gauge problem of TDCIS theory.
It was shown that the velocity gauge is size inconsistent, with respect to the number of active core orbitals, due to the lack of virtual transitions to doubles (D).
An effective potential was introduced  to account for non-resonant core polarization by usage of TRK theory.
The simulated photoelectrons from TDCIS-TRK theory were found to agree well with those from length gauge TDCIS, with a low number of angular momenta being required for numerical convergence of laser-assisted photoionization.
Formally, there is no justification to claim that gauge invariance is achieved, but the TDCIS-TRK theory seems to perform at a level comparable to the length gauge TDCIS, with reasonable energy shifts obtained for photoelectrons.
Contrary to previous attempts to achieve full gauge invariance in TDCIS between length and velocity gauge of TDCIS~\cite{satoGaugeInvariantAS2018}, the present correction can be implemented with negligible numerical cost in propagation with a spectral representation of the wavefunction, which may prove important for efficient implementations of TDCIS for more complex targets, such as molecules or heavy atoms.

Interestingly, the TDCIS-TRK simulations were essentially converged with the active core space limited to $\{2p\}$ in neon atoms at the considered field parameters.
While this may seem reasonable at first glance, because laser-assisted photoionization processes from $2p$ are known to be weakly correlated with $2s$ in the length gauge  \cite{dahlstromDiagrammaticPRA2012}, the present theory was derived from the velocity gauge, where the convergence of many-body effects in laser fields are much more subtle  \cite{vinbladhManybodyPRA2019}.
As an example, we showed that convergence of the effective number of active electrons: $\tilde{N}\rightarrow N$, was only reached for the {\it full} active core within RPAE (Table~\ref{tab:N-values}).

In summary, the TDCIS-TRK theory allows for a reduced number of angular momenta and an efficient frozen space in time-dependent simulations of laser-assisted photoionization.
The theory was used to predict a new type of dynamical interference phenomenon for atoms, under experimentally viable conditions, by separately controlling the flux and the ponderomotive shift by two-color fields.
In future works, the TDCIS-TRK theory may prove useful for studies of other observables, such as attosecond time-delays in photoionization with strongly-driven resonant transitions, HHG with multiple active channels, interferometric above-threshold ionization processes and photoionization dynamics by intense attosecond pulse trains generated by seeded FELs.

\appendix
\section{Numerical specifications}
Our calculations of laser-assisted photoionization are performed using the vector potential
\begin{equation}
  \label{eq:fields}
  A(t) = \left[A_{0,\Omega} \sin(\Omega t) + A_{0,\omega} \sin(\omega t)\right] f(t),
\end{equation}
with $\hslash \omega = \probefreqData{}\, \mathrm{eV}$ and $\hslash \Omega = \pumpfreqData{}\, \mathrm{eV}$.
To maintain the notion of a constant ponderomotive energy, while alleviating the spectral profile of the electric field, we use a flat-top pulse with smooth but sufficiently rapid truncation, given by
\begin{equation}
  \label{eq:flat-top}
  f(t) =
  \begin{cases}
    1, \quad & |t| \le \frac{\tau}{2} \\
    \exp\left[-\tan\left(\pi\frac{|t| - \frac{\tau}{2}}{t_{\mathrm{max}}-\tau}\right)^2\right], \quad & \frac{\tau}{2} < |t| \le \frac{t_{\mathrm{max}}}{2} \\
  0, & \textrm{otherwise.}
  \end{cases}
\end{equation}
This envelope is equal to unity for a width of $\tau$ and is smoothly suppressed between $|\tau|/2$ and $|t_{\mathrm{max}}|/2$. The width of the pulses is set to $\tau = \tauData{}\, \mathrm{fs}$ and the smooth suppression starts at $\tau_{\mathrm{off}} = \truncationLengthData{}\, \mathrm{fs}$.

In our last example, we however make use of more realistic truncated Gaussian pulses, adapted from Ref.~\cite{patchkovskiiSimpleCPC2016}.
In this case, the vector potential envelope is given by
\begin{widetext}
\begin{equation}
  \label{eq:gaussian}
  f(t) =
  \begin{cases}
    \exp[-\alpha t^2], \quad & |t| \le t_{\mathrm{off}} \\
  \exp\left[-\alpha \left(t_{\mathrm{f}} + \frac{2}{\pi}(t_{\mathrm{max}}-t_{\mathrm{f}}) \tan(\frac{\pi}{2}\frac{|t| - t_{\mathrm{f}}}{t_{\mathrm{max}} - t_{\mathrm{f}}})\right)^2\right], \quad & \frac{\tau}{2} < |t| \le t_{\mathrm{max}} \\
  0, & \textrm{otherwise}
  \end{cases},
\end{equation}
\end{widetext}
where $\alpha = 2\log(2)/\tau^2$ is chosen such that the pulse length, expressed as full width at half maximum, is specified for the intensity profile of the pulse.
This is valid in the long-wavelength limit.
The vector potential follows a Gaussian profile within the width, specified in terms of standard deviations of the intensity profile, $\sigma = T/(2\sqrt{2 \log{2}})$, of $|t| \leq t_{\mathrm{off}} = \toffTruncatedGaussian\sigma$, and is truncated at $t_{\mathrm{max}} = \tmaxTruncatedGaussian\sigma$.

To resolve the photoelectron spectra we use the time-dependent surface flux~\cite{taoPhotoelectronNJP2012,bertolinoPropensityJPBAMOP2020} (t-SURFF) and infinite-time surface flux~\cite{moralesISURFJPBAMOP2016,carlstromGeneral2022} (iSURF) methods.
A limiting assumption of t-SURFF is that the total wavefunction be separable into a bound ionic part and a free electronic part.
This assumption breaks due to the long-range Coulomb potential, which scales with the inverse of the radial distance in space. However, the photoelectrons are recorded at $R_0=\tsurffRData$, where the effect of the Coulomb potential is sufficiently small to not affect our results.

\begin{acknowledgements}
JMD acknowledges support from the Swedish Research Council: 2018-03845, the Olle Engkvist Foundation: 194-0734 and the Knut and Alice Wallenberg Foundation: 2017.0104 and 2019.0154.
We thank Andrea Idini for fruitful discussions about many-body treatment, and Per Eng-Johnsson and Anne L'Huillier for fruitful discussions about laser-assisted dynamical interference.
\end{acknowledgements}

\bibliographystyle{apsrev4-2}
\bibliography{polarization2020}
\end{document}